\documentclass[12pt,a4paper,twoside]{article}
\usepackage{epsfig}
\usepackage{amssymb,latexsym,amsfonts}

\newcommand{\be}{\begin{equation}}
\newcommand{\ee}{\end{equation}}
\newcommand{\bd}{\begin{displaymath}}
\newcommand{\ed}{\end{displaymath}}
\newcommand{\ba}{\begin{array}}
\newcommand{\ea}{\end{array}}
\newcommand{\bt}{\begin{tabular}}
\newcommand{\et}{\end{tabular}}

\newcommand{\bea}{\begin{eqnarray}}
\newcommand{\eea}{\end{eqnarray}}
\newcommand{\bean}{\begin{eqnarray*}}
\newcommand{\eean}{\end{eqnarray*}}
\newcommand{\non}{\nonumber}

\newcommand{\hlf}{\frac{1}{2}}

\newcommand{\inp}[2]{\langle #1, #2 \rangle}

\newcommand{\Z}{\mathbb{Z}}

\newcommand{\R}{\mathbb{R}}
\newcommand{\C}{\mathbb{C}}

\begin{document}
\begin{titlepage}

\begin{center} 
\today \hfill                  hep-th/0309106

\vskip 2 cm
{\Large \bf The topology of U-duality (sub-)groups\\}
\vskip 1.25 cm
{Arjan Keurentjes\footnote{email address: Arjan@tena4.vub.ac.be}}\\
\vskip 0.5cm
{\sl Theoretische Natuurkunde, Vrije Universiteit Brussel,\\ Pleinlaan
  2, B-1050 Brussels, Belgium \\}
\end{center}
\vskip 2 cm
\begin{abstract}
\baselineskip=18pt
We discuss the topology of the symmetry groups appearing in
compactified (super-)gravity, and discuss two applications. First, we
demonstrate that for 3 dimensional sigma models on a symmetric space
$G/H$ with $G$ non-compact and $H$ the maximal compact subgroup of
$G$, the possibility of oxidation to a higher dimensional theory can
immediately be deduced from the topology of $H$. Second, by comparing
the actual symmetry groups appearing in maximal supergravities with
the subgroups of $SL(32,\R)$ and $Spin(32)$, we argue that these
groups cannot serve as a local symmetry group for M-theory in a
formulation of de Wit-Nicolai type.  
\end{abstract}

\end{titlepage}
\section{Introduction}

Since the construction of supergravity theories, and the discovery of the
Cremmer-Julia groups of compactified 11 dimensional supergravity
\cite{Cremmer:1978km, Cremmer:1979up}, it
has been clear that Lie groups and algebra's play an important
role in this field. However, most of the attention is confined to the subject of Lie algebra's. In this
paper we will study the topology of some of the (sub-)groups present
in (super-)gravity, and hope to convince the reader that examination
of these global properties leads to relevant information about the theory.

In section \ref{topo} we give some technical background, and quote a
useful theorem. Then we apply the methods to two different, but related
topics: The theory of oxidation, and a recent proposal for a ``generalized holonomy''-group as a
symmetry of 11 dimensional supergravity and M-theory. 

Dimensional reduction of a theory leads to a lower dimensional
theory. Oxidation is the inverse to this process, the reconstruction
of a higher dimensional theory from the lower dimensional one. One of
the most interesting cases to consider is a 3 dimensional sigma model on a coset $G/H$, coupled to gravity
\cite{Julia:1980gr, Julia:1982gx, Breitenlohner:1987dg,
  Cremmer:1999du}. In \cite{Keurentjes:2002xc}, it
was demonstrated how the group $G$ encodes all the higher
dimensional theories that lead to a $G/H$ coset upon dimensional
reduction to 3 dimensions, including the possibility of inequivalent
dual theories (see \cite{Henry-Labordere:2002dk} for an alternative
approach, covering truncations of maximal supergravity). In section
\ref{oxi} of this paper, we look at the maximal compact subgroup $H$,
and argue that the structure of this group is crucial to the inclusion
of fermionic representations. As a consequence, the topology of $H$
inside $G$ immediately gives a criterion for oxidizability (even
without further study of the group $G$). 

A second topic will be the study of proposals for symmetry groups
of M-theory \cite{Duff:2003ec, Hull:2003mf} (see
\cite{Papadopoulos:2003pf, West:2003fc} for subsequent work). An
important concept in the study of backgrounds for M-theory is the
holonomy (of the spin bundle) on the (geometric) background; it
determines the amount of preserved supersymmetries, and many
properties of the low-dimensional theory. However, a background may
not be ``purely'' geometrical, but instead include fluxes. In
compactification, the fields giving rise to these fluxes merge
together with geometric degrees of freedom and are mixed by (subgroups
of) the Cremmer-Julia groups. The existence of formulations of 11
dimensional supergravity with the compact subgroup of a Cremmer-Julia
group appearing as a local symmetry\footnote{Strictly speaking it is
  the universal covering of the compact subgroup of the Cremmer-Julia
  group, as we will explain in sections \ref{oxi} and \ref{holo}.}
\cite{deWit:1985iy}, gives rise to the ``generalized
holonomy''-conjecture: There exists a suitable extension of the
structure group on the spin bundle, such that (in a suitable
formulation) a generic background can be described as having
``generalized holonomy'', that is holonomy in
this extended structure group \cite{Duff:2003ec}. 

The immediate question is whether 11 dimensional supergravity allows a
formulation, such that all possible local symmetries from space-,
time- or null-like reduction can be realized within some group, and what this structure
group should be; because it represents a \emph{local} symmetry it seems unavoidable that
it must be a symmetry of some formulation of the non-perturbative
extension of 11 dimensional supergravity, M-theory. In
\cite{Hull:2003mf} it is proposed that the structure group may be a
simple finite dimensional group; based on
the requirements that such a group must meet, and an analysis of
symmetries generated by the 11 dimensional Clifford algebra,
$SL(32,\R)$ is put forward as a candidate.

In section \ref{holo} we examine the topology of the symmetry groups
appearing in maximal supergravities, and find that the group $SL(32,\R)$ does
not contain them all. Similarly, the group $Spin(32)$ proposed
in \cite{Duff:2003ec} also appears problematic. We conclude that these
groups cannot be straightforwardly promoted to symmetry groups of M-theory.
  
\section{The topology of (sub-)groups} \label{topo}

Although often neglected in the physics literature, the topology
of (sub-)groups is relevant to a variety of effects. Elements of the
discussion here were crucial in the discovery of a new class of vacua
for Yang-Mills theory on a 3-torus \cite{Keurentjes99, Kac,
  Borel}. An application to string and M-theories leads to new string-
and M-theory vacua \cite{deBoer:2001px}. The first part of our
discussion follows \cite{Kac}, similar results can be found in
\cite{Borel}. 

The groups we will be discussing are compact groups. A central theorem
in Lie-group theory says that every compact simple Lie-group $H$ has a
universal covering $\widetilde{H}$, which is simply connected. The
fundamental group of $H$ is given by a subgroup $\pi_1(H)=Z$ of the
center of $\widetilde{H}$, and the group $H$ is isomorphic to
$\widetilde{H}/Z$. 

Therefore, to deduce the fundamental group of $H$, it suffices to
compare the center of $H$ with the center of $\widetilde{H}$. In practice,
we are not dealing with abstract groups, but with representations
of $H$. These are always representations of $\widetilde{H}$, and it will
be necessary to check whether the elements of the center of $\widetilde{H}$
are realized non-trivially on these representations. The different
topologies that can be realized are enumerated by the possible
different subgroups of $Z$. 

At the level of the lattices associated to the Lie algebra $h$ that
generates $H$, this has a nice realization \cite{Kac}. Let
$Q^{\vee}(H)$ be the coroot-lattice of $h$, and let $P^{\vee}(H)$
be the coweight lattice of $h$ 
\be
P^{\vee}(H) = \{ \zeta \in h | \exp (2 \pi i \zeta) = 1 \}
\ee
Then the fundamental group of $H$ is given by
\be \label{cowcor}
\pi_1(H) = P^{\vee}(H)/Q^{\vee}(H)
\ee
 
The groups we are interested in are subgroups of larger groups. We
quote a useful theorem from \cite{Kac}, which is devoted to
subgroups of compact groups (similar results can be found in
\cite{Borel}): It summarizes many computations that can also be done
case by case \cite{Keurentjes99}. 

Let $r$ be the rank of the algebra
$h$, and $\alpha_i$ ($1 \leq i \leq r$) be a set of simple roots for $h$. Define the
fundamental coweights by $\inp{\alpha_i}{\omega_j} = \delta_{ij}$
(where $\inp{}{}$ is the bilinear Killing form on $h$). The
root integers $a_i$ (resp. coroot integers $a_i^{\vee}$) are given by
expanding the highest root $\theta$ (resp. its coroot $\theta^{\vee}= 2 \theta/
\inp{\theta}{\theta}= \theta$)  in the simple roots (coroots): 
\be
\theta = \sum_{j=1}^r a_j \alpha_j = \sum_{j=1}^r a_j^{\vee}\frac{2
  \alpha_j}{\inp{\alpha_j}{\alpha_j}} 
\ee
Furthermore, one introduces $a_0= a_0^{\vee}=1$.

Then one has the following \\
{\bf Theorem 1} (\cite{Kac} theorem 1b/c) Let $H$ be a connected simply
connected (almost) simple Lie group. Consider the group element
\be \label{elem}
\sigma_{\vec{s}} = \exp \{ 2 \pi i \sum_{j=1}^r s_j \omega_j \}
\ee
where
\be \label{constr}
\vec{s} = (s_0,s_1, \ldots s_r), \quad s_j \in \R \quad, s_j \geq 0 , \quad \sum_{j=0}^{r} a_j s_j
=1.
\ee
The centralizer of $\sigma_{\vec{s}}$ (in $H$) is a connected compact
Lie group which is a product of $U(1)^{n-1}$, where $n$ is the number
of non-zero $s_j$, and a connected semi-simple group whose Dynkin diagram is obtained from the extended
Dynkin diagram of $H$  by removing nodes $i$ for which $s_i \neq 0$. Moreover,
the fundamental group of the centralizer of $\sigma_{\vec{s}}$ is
isomorphic to a direct product of $\Z^{n-1}$ and a cyclic group of
order $a_{\vec{s}}$ where 
\be
a_{\vec{s}} = \gcd \{ a_j^{\vee} | s_j \neq 0 \}
\ee

{\bf Proof} See \cite{Kac} and references therein. Note the different
ranges appearing in the sums in equations (\ref{elem}) and (\ref{constr}). 

\begin{figure}[ht]
\begin{center}
\includegraphics[bb=100 500 500 800, width=10cm]{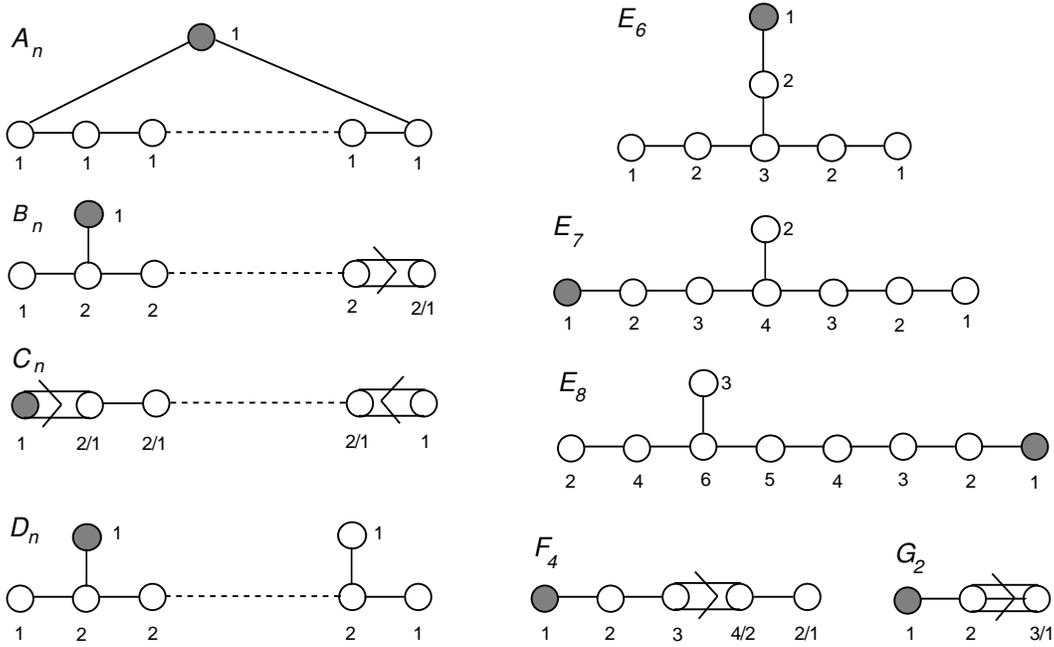}
\caption{Extended Dynkin diagrams, root and coroot integers.}\label{fig} 
\end{center}
\end{figure}

We have depicted the extended Dynkin diagrams, root and coroot
integers in figure \ref{fig}. Extended nodes are shaded, omitting them
leads to the standard Dynkin diagram. In case root and coroot integers differ,
we have denoted them as $a_j/a_j^{\vee}$.

Theorem 1 allows an easy computation of the topology of the compact
subgroup $H$ of a non-compact group $G$ that is generated by a Cartan
involution that is inner. Note that if exactly one of the
$s_{j \neq 0}=\hlf$, then $\sigma_{\vec s}$ squares to 1 (acting in
the adjoint representation), and can be used as a Cartan involution. This generates a non-compact
real form $G$, and its maximal compact subgroup $H$ is the
centralizer of $\sigma_{\vec s}$, which is easily found by using
theorem 1. There are 3 ways to arrange this, and meet the other constraints.

First, suppose the node $j$ corresponds to a long root, with
  $a_j=2$. This situation occurs in all extended $B,D,E,F,G$ diagrams
  (see figure \ref{fig}). Necessarily, $a_j^{\vee}=2$ and theorem 1
  tells us that the fundamental group of $H$ is $\pi_1(H) = \Z_2$. 

Second, suppose that the node $j$ represents a short root with 
$a_j=2$.  This situation occurs in the $B,C$ and $F$ diagrams. Now
$a_j^{\vee}=1$, and the fundamental group of the maximal compact
subgroup is given by $\pi_1(H) = 0$

Third, suppose that the node $j$ represents a long root with
$a_j=1$. Then we solve (\ref{constr}) by setting $s_0 = s_j =
\hlf$. This can be done for all groups except $E_8$, $F_4$ and
$G_2$. As $a_0^{\vee}=a_j^{\vee}=1$, theorem 1 tells us that the
maximal compact subgroup has $\pi_1(H) = \Z$. 
 
These results cover all cases where the Cartan involution is an inner
automorphism (see table \ref{tab1}, this can also easily be proven by
invoking Theorem 1a from \cite{Kac}). In the remaining cases, the
Cartan involution is an \emph{outer} automorphism. We know no general
theorem applicable to this situation, and will compute the
fundamental group case by case instead. We do
not have to compute the fundamental group for all (infinitely many)
representations $H$, but can restrict to a few well chosen
irreducible representations (irreps), as we will know explain.

By a simple rewriting of (\ref{cowcor}), the fundamental group of $H$ is also given by
\be
\pi_1(H) = P(H)/Q(H)
\ee
where $P$ is the lattice generated by the
weights of the representations present in the theory, and $Q$ is the
root lattice. Consider the \emph{full} weight lattice of
$\widetilde{H}$, which is the dual lattice to the coroot lattice,
and includes representations that may not be present
in our theory of interest. The root lattice is a sublattice of the full
weight lattice, and quotienting the full weight lattice by the root
lattice introduces a grading on the weight lattice. Because in an
irrep, the various weights differ by a root, all weights of one irrep
belong to a single equivalence class, called a congruence
class. The lattice $P$ is a sublattice of the full weight lattice, and only a subset of the
possible congruence classes may be present. It then suffices to check
the decomposition into subgroups for one representative of each
congruence class present. Better still, it is sufficient to restrict to
those congruence classes that \emph{generate} all the other
classes. This reduces the problem to a finite computation. It gives
the full answer, because the non-trivial realization of the elements
of the center of the $\widetilde{H}$, which gives the fundamental
group, depends only on the congruence class. 
 
\section{Oxidation and the maximal compact subgroup} \label{oxi}

In this section we apply the math of section \ref{topo} to the theory of
oxidation. Consider a 3 dimensional sigma model on a coset $G/H$,
coupled to gravity. Is it possible to interpret this as the effective
theory resulting from toroidal compactification of a higher
dimensional theory? The answer is often yes, as demonstrated by the
explicit dimensional reduction of higher dimensional theories
\cite{Julia:1980gr, Julia:1982gx, Breitenlohner:1987dg,
  Cremmer:1999du}. In \cite{Keurentjes:2002xc} we gave a constructive
approach, in which one can derive the existence of the higher
dimensional theory, and the details (such as field content, dynamical
and constraint equations) from an analysis
of the properties of $G$. Here instead, we focus on the maximal
compact subgroup $H$, and show that its topology immediately tells us
whether the 3 dimensional coset model on $G/H$ can be oxidized. 

To motivate our discussion, we consider the possibility of adding
fermions to the theory (the papers \cite{Julia:1980gr, Julia:1982gx,
  Breitenlohner:1987dg, Cremmer:1999du,Keurentjes:2002xc,
  Henry-Labordere:2002dk} focus almost exclusively on bosons). 
Consider the reduction of General Relativity from $d$ to 3 dimensions
\cite{Cremmer:1997ct}. This gives rise to a sigma model on
$SL(d-2)/SO(d-2)$. The $SO(d-2)$ group appearing here can be
thought of as the remnant of the helicity group for massless fields
\cite{Julia:1980gr}. The actual maximal compact subgroup in $SL(n,\R)$ is $SO(n)$ (see appendix
\ref{app}), and $\pi_1(SO(n)) = \Z_2$ (for $n>2$), which is
of course crucially related to fermions. If we want to add massless fermions to
the theory, these necessarily transform in representations of the
helicity group $Spin(d-2)$ that are not representations of
$SO(d-2)$. In the special case $d=4$ we have $SO(2)$, and should
normalize the unit of charge. It is customary to normalize these in
such a way that the bosons in the theory have integral charge
(spin). Then the fermions turn out to have half-integral spins, so also
here we are dealing with a two-fold cover of the group relevant to the
bosons.

In \cite{Keurentjes:2002xc} it was argued that the index 1 embeddings of
$SL(d-2,\R) \subset G$ give all the relevant information to
reconstruct the higher dimensional theories. Via the embedding of
$H \subset G$, and the embedding of $SO(d-2) \subset SL(d-2,\R)
\subset G$, the representations of the helicity group are encoded in $H$. 

But $SO(d-2)$ is also relevant to the existence of fermions. The fact
that these transform in the double cover of the group remains true
after dimensional reduction. But then it is crucial that \emph{$H$
  must have a topology that is compatible with that of
  $SO(d-2)$}. Roughly, the $2 \pi$ rotation that leaves bosons
invariant and multiplies fermions with a minus sign, must be realized
in the local symmetry groups $H$, in the compactified theory. More precise, for a theory
that has its origin in $d$ dimensions with $d>4$, the topology of $H$ has to be
such that it can accommodate the $\Z_2= \pi_1(SO(d-2))$, otherwise one
can violate spin-statistics via compactification. The same is true in
$d=4$; the $SO(2)$ helicity group \emph{must} allow a double cover,
which is necessarily also an $SO(2)$. Hence if the sigma model on
$G/H$ can be oxidized to $d$ dimensions, we expect
\be \label{neces}
\pi_1(H) \supset \Z \  (\textrm{for }d=4), \qquad \pi_1(H) \supset
\Z_2 \ (\textrm{for }d>4).
\ee

\begin{table}
\begin{center}
\begin{tabular}{|c|c|c||c|c|}
\hline
$g$ & $h$ & Cartan &  $\pi_1(H)$ & $d$ \\
\hline
\hline
$sl(2)$ & $u(1)$ & in & $\Z$ & 4 \\
$sl(n > 2)$ & $so(n)$ & out & $\Z_2$ & $n+2$ \\
\hline
$su^*(2n)$ & $sp(n)$ & out & 0 & 3 \\
\hline
$su(p,q)$ & $su(p) \oplus su(q) \oplus u(1)$ & in & $\Z$ & 4 \\
\hline
\hline
$so(1,2) \cong sl(2)$ & $u(1)$ & in & $\Z$ & 4 \\
$so(1,n)$, $n>2$ & $so(n)$ & in/out & 0 & 3 \\
$so(2,n)$, $n>2$ & $u(1) \oplus so(n)$ & in & $\Z$ & 4 \\
$so(p,q)$, $p,q >2$ & $so(p) \oplus so(q)$ & in/out & $\Z_2$ & max($p+2,q+2$),\\
 & & & & 6 (iff. $p$ or $q$ $\geq$ 4)\\
\hline
$so^*(2n)$ & $u(1) \oplus su(n)$ & in & $\Z$ & 4 \\
\hline
\hline
$sp(n,\R)$ & $u(1) \oplus su(n)$ & in & $\Z$ & 4 \\
\hline
$sp(p,q)$ & $sp(p) \oplus sp(q)$ & in & 0 & 3 \\
\hline
\hline
$e_{6(6)}$ & $sp(4)$ & out & $\Z_2$ & 8 \\
\hline
$e_{6(2)}$ & $su(6) \oplus su(2)$ & in & $\Z_2$ & 6 \\
\hline
$e_{6(-14)}$ & $so(10) \oplus u(1)$ & in & $\Z$ & 4 \\
\hline
$e_{6(-26)}$ & $f_{4}$ & out & 0 & 3 \\
\hline
\hline
$e_{7(7)}$ & $su(8)$ & in & $\Z_2$ & 10,8 \\
\hline
$e_{7(-5)}$ & $so(12) \oplus su(2)$ & in & $\Z_2$ & 6 \\
\hline
$e_{7(-25)}$ & $e_6 \oplus u(1)$ & in & $\Z$ & 4 \\
\hline
\hline
$e_{8(8)}$ & $so(16)$ & in & $\Z_2$ & 11,10 \\
\hline
$e_{8(-24)}$ & $e_7 \oplus su(2)$ & in & $\Z_2$ & 6 \\
\hline
\hline
$f_{4(4)}$ & $sp(3) \oplus su(2)$ & in & $\Z_2$ & 6 \\
\hline
$f_{4(-20)}$ & $so(9)$ & in & $0$ & 3 \\
\hline
\hline
$g_{2(2)}$ & $su(2) \oplus su(2)$ & in & $\Z_2$ & 5 \\
\hline 
\end{tabular}
\caption{Non-compact algebra's $g$, maximal compact subalgebra's $h$, Cartan
  involution, the fundamental group of
  $H$, and the maximal oxidation dimension
  $d$.} \label{tab1}
\end{center}
\end{table}
 
This gives a \emph{necessary} criterion. We now study the
possible symmetric spaces $G/H$ with simple non-compact $G$, and the topology of
the maximal compact subgroup $H$. The results are given in table
\ref{tab1}. The list of symmetric spaces is taken from
\cite{Helgason}. We have listed whether the non-compact form is
generated by an inner or outer (Cartan) involution. The
topologies of compact subgroups were computed with theorem 1 of
section \ref{topo} (for inner Cartan involutions), and in appendix
\ref{app} (for outer Cartan involutions). The maximal oxidation
dimension for various theories can be found in
\cite{Keurentjes:2002xc}, and references therein. Multiple entries
indicate branches with different end-points.   

Interestingly, restricting to \emph{simple} Lie
groups, the necessary criterion (\ref{neces}) turns out to be a
\emph{sufficient} criterion; the subset-symbols can be changed to
equal signs! Hence we have:

{\bf Theorem 2:} Consider a sigma model in 3 dimensions on a symmetric
space $G/H$, with $G$ a simple non-compact group and $H$ its maximal compact
subgroup, coupled to gravity. This sigma model can be oxidized to a
higher dimensional model if and only if the group $H$, as embedded in $G$, is not simply
connected. Moreover, the maximal oxidation dimension $d$ is given by:
\be 
\ba{lll}
d=3 & \textrm{if} & \pi_1(H) = 0 \\
d=4 & \textrm{if} & \pi_1(H) = \Z \\
d > 4 &  \textrm{if} & \pi_1(H) = \Z_2 \\
\ea 
\ee

{\bf Proof:} By inspection of table \ref{tab1}.

This can be immediately extended to non-simple groups. When the
symmetric space is of the form $\prod_i G_i/H_i$, the compact subgroup
has fundamental group $\prod_i\pi_1(H_i)$, and one can choose to
oxidize from any simple factor, to which the above criterion can
be applied. 

The complex groups are generated from two copies of the same algebra, with
the outer automorphism that exchanges the two copies as Cartan
involution. This gives a symmetric space $H^{\C}/H$, where $H$ is the
simply connected compact form of the group. Hence criterion
(\ref{neces}) suggests they cannot be oxidized, which is indeed true
\cite{Keurentjes:2002xc}.
 
We conclude with a remark relevant to oxidation of theories with
fermions. In \cite{Keurentjes:2002xc} we have explained how the decomposition of
the non-compact group $G$ into subgroups leads to information on the
bosonic content of the higher dimensional theories. As the fermions do
not arise in representations of $G$, they were not considered. When
one instead turns to the compact groups $H$ and $\widetilde{H}$, it
remains true that $Spin(d-2)$ should play a role as helicity group, and
therefore the decompositions of $H$ into $Spin(d-2)$ are relevant.

A well known example is the embedding $Spin(16) \rightarrow Spin(9)$
that is relevant for retrieving the 11 dimensional supergravity from
maximal 3 dimensional supergravity. The bosonic degrees of freedom of
the 3-d theory are organized in the $\mathbf{128_s}$ of
$Spin(16)$, while the fermions are in the
other spin irrep $\mathbf{128_c}$ \cite{Marcus:1983hb}. Decomposing to
$Spin(9)$ this gives
\be
\mathbf{128_s} \rightarrow \mathbf{44 \oplus 84} \qquad
\mathbf{128_c} \rightarrow  \mathbf{128}
\ee
Actually, all irreps of $Spin(16)/\Z_2 \subset E_{8(8)}$
(which are bosons, like the $\mathbf{128_s}$) decompose to irreps of
$SO(9)$.  Similar decompositions can be made for $Spin(16) \rightarrow
Spin(d-2) \times \widetilde{H}_d$ for maximal supergravity, but also
for the other oxidizable theories in table \ref{tab1}. This seems to
hint at a generalization of the concept of ``spin structure'' with
other groups than orthogonal ones.

\section{Holonomy and symmetries of M-theory} \label{holo}

Before investigating the proposal of \cite{Hull:2003mf}, let us
reexamine its logic. 

Upon compactification of 11 dimensional supergravity, there exist
formulations of the resulting effective theories exhibiting a symmetry
$Spin(1,d-1) \times \widetilde{H}_d$, where the first factor is the
Lorentz-group, and $\widetilde{H}_d$ is (the double cover of) the
maximal compact subgroup of a Cremmer-Julia group
\cite{Cremmer:1979up, Cremmer:1997ct}. The existence of these ``hidden symmetries''
prompts the question whether they are a consequence of
compactification, or are already present in some form in the higher
dimensional theory. An answer to this question was given in
\cite{deWit:1985iy}, where formulations of 11 dimensional supergravity
with a local $Spin(1,d-1) \times \widetilde{H}_d$ invariance were constructed.  

Now because $\widetilde{H}_d$ is a \emph{local} gauge symmetry in such
formulations, it is hard to conceive of a way of breaking it. In
compactification it is ``broken'' by the presence of boundary
conditions. More accurately, this means that there is non-trivial
\emph{holonomy} in the group $\widetilde{H}_d$, such that it is no longer a manifest
symmetry of the lower dimensional theory. The group $\widetilde{H}_d$ includes the
Lorentz group $Spin(11-d)$ of the compactified dimensions, but as the
Cremmer-Julia analysis shows, it is almost always bigger. This allows the
possibility of a background with holonomy in (a subgroup of)
$Spin(1,d-1) \times \widetilde{H}_d$ \cite{Duff:2003ec}. 

It is now proposed to generalize this concept to a new formulation of
11 dimensional supergravity, with a symmetry group $\widetilde{H}_0$ which
contains all possible $\widetilde{H}_d$ \cite{Hull:2003mf}. Clearly,
$\widetilde{H}_0$, should include all groups $Spin(1,d-1)\times
\widetilde{H_d}$ from space-like compactifications, as well as analogous groups for time-like, and
null-compactifications. In \cite{Hull:2003mf} it is claimed that the
finite-dimensional group $SL(32,\R)$ meets all the required
properties. 

We will here re-examine this claim. The works \cite{Duff:2003ec,
  Hull:2003mf, Papadopoulos:2003pf} study the symmetry at the level of
  the supercovariant derivative, acting on spinors. These spinors form the
  supersymmetry parameters, and their $\widetilde{H}_d$ representation carries
  over to their gauge fields, the gravitini. In this study of the
  global properties of the symmetry groups, we also have to look at
  their realization on the other fields, organized in (other) $\widetilde{H}_d$ 
  representations.  

\begin{table}
\begin{center}
\begin{tabular}{|c|cc|}
\hline
d & $G_d$ & $H_d$ \\
\hline
11 & $\{e\}$ & $\{e\}$ \\
10 & $\R$, $SL(2,\R)$  & $\{e\}, SO(2)$  \\
9  & $SL(2,\R) \times \R$ & $SO(2)$ \\
8  & $SL(3,\R) \times SL(2,\R)$ & $SO(3) \times SO(2)$ \\
7  & $SL(5,\R)$ & $SO(5)$ \\
6  & $Spin(5,5)$ & $(Sp(2) \times Sp(2))/\Z_2$ \\
5  & $E_{6(6)}$ & $Sp(4)/\Z_2$ \\
4  & $E_{7(7)}$ & $SU(8)/\Z_2$ \\
3  & $E_{8(8)}$ & $Spin(16)/\Z_2$ \\
\hline
\hline
2  &            & $Spin(16) \times Spin(16)$ \\
1  &            & $Spin(32)$               \\
0  &            & $SL(32,\R)$            \\
\hline    
\end{tabular}
\caption{Above the double line: Cremmer-Julia groups $G_d$; their compact
  subgroups $H_d$. Below the double line: Candidate ``generalized
  holonomy''-groups in lower dimensions} \label{tab2} 
\end{center}
\end{table}

In table \ref{tab2} we have listed the Cremmer-Julia groups $G_d$ and
their maximal compact subgroup $H_d$ that appear in space-like
reduction. We will not discuss the groups for time-like
reductions\footnote{The table for these groups would look similar to
  table \ref{tab2},with the groups $H_d$ replaced by the appropriate (quotients
  of covering-)groups of \cite{Hull:1998br}, but would be incomplete
  without a discussion of the possibility of disconnected components
  of the group, which is outside the scope of this paper.}
\cite{Hull:1998br}, and null reductions \cite{Duff:2003ec}. We have
  taken special care to list the subgroups with their proper
  topologies. Below the double line, we have listed the candidate
  ``generalized holonomy''-groups, \emph{without} making claims on
  their topology: Until further notice, we remain open-minded on what
  these are subgroups of, or what they are acting on, and as such, it
  is not yet clear what their topology should be. 

We should emphasize that the groups $H_d$ listed are subgroups of the
Cremmer-Julia groups, and as such only refer to the \emph{bosonic}
sectors of the theory. The fermions in the theory are \emph{not} in
representations of $H_d$; instead they form representations of a
double cover of $H_d$, $\widetilde{H}_d$. This sharpens the well-known
fact that the fermions do not form representations of $G_d$; actually,
\emph{there exist no representations of $G_d$ that include the
  representations of the fermions in their $H_d$
  decompositions}. Hence, no additional auxiliary fields can ever cure
this, and the symmetry groups $G_d$ can represent at most symmetries of the
bosonic sector of the theory, \emph{never} of the full theory. 

Therefore, the situation is much like in standard General Relativity
  where fermions do not form representations of $SO(1,d-1)$, but only
  of its simply connected cover $Spin(1,d-1)$. When fermions are
  included we have to use a formalism with vielbeins $\in GL(d,
  \R)/SO(1,d-1)$, local frames, and gamma matrices. Similar concepts
  should be invoked in the proposed formulation of 11 dimensional
  supergravity/M-theory.  

We discuss some theories from table \ref{tab2} in somewhat
more detail, to give the reader an impression what these statements on
topology actually mean.
\begin{itemize}
\item In 6 dimensions, the scalars are in the $\mathbf{(5,5)}$ of
  $Sp(2) \times Sp(2)$, there are vectors in the $\mathbf{(4,4)}$,
  while the antisymmetric tensors (self-dual and anti self-dual)
  transform as $\mathbf{(5,1) \oplus (1,5)}$. None of these
  representations realizes the diagonal $\Z_2$ in the $\Z_2 \times
  \Z_2$ center of $Sp(2) \times Sp(2)$, so they are all irreps of
  $(Sp(2) \times Sp(2))/\Z_2$. The gravitini can be split in parts
  with opposite chirality; they transform in $\mathbf{(4,1) \oplus
  (1,4)}$. The remaining fermions transform as $\mathbf{(5,4) \oplus
  (4,5)}$. The fermionic irreps realize all center elements, hence
  the total symmetry of the theory is $Sp(2) \times Sp(2)$.
\item In 5 dimensions \cite{Cremmer:1980gs} the scalars transform in the $\mathbf{42}$, and
  the vectors in the $\mathbf{27}$. Neither of the 2 realizes the
  $\Z_2$ center of $Sp(4)$, hence they are irreps of $Sp(4)/\Z_2$. The
  gravitini are in the $\mathbf{8}$, while the other fermions are in
  the $\mathbf{48}$. These realize the full center, and the full
  symmetry is $Sp(4)$. 
\item In 4 dimensions \cite{Cremmer:1979up}, the scalars are in the $\mathbf{70}$ of $SU(8)$
  (the antisymmetric 4-tensor), the vectors in the $\mathbf{28}$ (the antisymmetric 2
  tensor); each of these is an irrep of $SU(8)/\Z_2$. The gravitini
  are in the $\mathbf{8}$, the other fermions in the $\mathbf{56}$,
  which are not irreps of $SU(8)/\Z_2$, but of $SU(8)$ .
\item In 3 dimensions \cite{Marcus:1983hb}, the compact subgroup of $E_{8(8)}$ is
  $Spin(16)/\Z_2$; the $\Z_2$ in the quotient is \emph{not} the one
  that leads to $SO(16)$, but one of the other $\Z_2$
  generators\footnote{The $Spin(32)/\Z_2$ group appearing in heterotic
  string theory is defined in the same way.} in the
  $\Z_2 \times \Z_2$ center of $Spin(16)$. The scalars are in the
  $\mathbf{128_s}$, which is the only spin irrep of $Spin(16)/\Z_2$;
  the gravitini are in the $\mathbf{16}$, and the fermions are in the
  $\mathbf{128_c}$ (the \emph{other} spin irrep), which are not irreps
  of $Spin(16)/\Z_2$. The full symmetry is therefore $Spin(16)$.
\end{itemize}

The reason to highlight these examples is that they are of direct
relevance to the formulations of 11 dimensional supergravity proposed
in \cite{deWit:1985iy}, that form an important ingredient in the
``generalized holonomy''-proposal. All these examples feature bosons
transforming in representations of an $H_d$ which is not simply
connected. The fermions transform in irreps of the simply connected
cover $\widetilde{H}_d$ of $H_d$, making the actual symmetry group a simply connected one.

The next element in the analysis is restriction to higher dimensional
groups. Clearly, each $H_d$ ($\widetilde{H}_d$) must contain all higher dimensional
entries $H_{d+n}$ ($\widetilde{H}_{d+n}$), leading to a chain
\be \label{embed}
Spin(16) \supset SU(8) \supset Sp(4) \supset Sp(2)\times Sp(2) \supset \ldots 
\ee 
As the reader can verify, in this chain we have mentioned the correct
topologies, e.g. the subgroup of $Spin(16)$ with an $su(8)$ algebra is
really $SU(8)$, and not some non-simply connected version. There
exists also a chain where all groups in (\ref{embed}) are quotiented
by a $\Z_2$, that describes the embeddings of the bosons only.  

We expect that this line of argument can be extended to the lower dimensional
part of the chain, the hypothetical ``generalized holonomy'' groups; even though
we do not know their explicit realization, they have to contain the
higher dimensional groups (the tildes can be omitted for the bosons)
\be
\widetilde{H}_d \supset \widetilde{H}_{d+1} \supset \widetilde{H}_{d+2} \supset \ldots.
\ee

Hence we turn to $\widetilde{H}_0 = SL(32,\R)$. Looking one entry higher in table \ref{tab2}, we
should require that it contains a group with algebra $so(32)$. Now
$SL(32,\R)$ has as maximal compact subgroup $SO(32)$, so there is only
one possible embedding. But, this compact group is
really $SO(32)$, and \emph{not} $Spin(32)$ (see appendix \ref{app}); there is no decomposition
of any $SL(32,\R)$ irrep that will ever give a spin irrep of $Spin(32)$.

It is not necessary to spend much time on the other decompositions;
$SO(32)$ decomposes in $SO(16) \times SO(16)$, and this group has no
$Spin(16)$ subgroups, only $SO(16)$ can be a subgroup. Irrespective of
the way in which the hypothetical $SL(32,\R)$ symmetry is realized,
there does not exist an $SL(32,\R)$ irrep that has the representation of the scalars and the
fermions in 3 dimensional supergravity in its decomposition (since these are in irreps of
$Spin(16)$, but not of $SO(16)$). Hence we conclude that $SL(32,\R)$
cannot be the final answer for a symmetry group of M-theory. 

The above argument relies on the embedding
\be
SL(32,\R) \supset SO(32) \supset (SO(16) \times SO(16)) \supset SO(16)
\ee
as the motivation for the original introduction of
$SL(32,\R)$ did \cite{Hull:2003mf}. Although this would be less attractive, one
might hope that another embedding chain  $SL(32,\R) \supset \ldots \supset Spin(16)$ with other
intermediate groups exists, but this is not so. This can
be argued by noticing that $SL(32,\R)$ has no index 1 orthogonal
subgroups; every orthogonal subgroup has to be index 2 or higher. But
every index 2 orthogonal subgroup is an $SO(n)$ and never a
$Spin(n)$ group. Yet higher index subgroups can be $Spin$-groups, for
example, $SO(16) \subset SL(16)$ has a $Spin(9)$ subgroup for which
the $\mathbf{16}$ dimensional vector irrep of $SO(16)$ decomposes into
the spin irrep of $Spin(9)$ of equal dimension. We however require
$\mathbf{128}$-dimensional irreps, and hence would need at least
$SO(128)$ for this type of construction to work. It therefore seems
impossible that $Spin(16)$ is a subgroup of $SL(32,\R)$. 

Another possibility that could save (part of) the proposal is, that
the realization of $SL(32,\R)$ on the theory is such that the
method of restriction to higher dimensional $H_d$ is invalid, but this
also seems an unlikely, and highly unattractive option.

We now turn to the proposed generalized holonomy group
$\widetilde{H}_1 = Spin(32)$. It leads to similar problems as
$\widetilde{H}_0 = SL(32,\R)$. The $so(16) \oplus so(16)$-subalgebra
of $so(32)$ generates $(Spin(16) \times Spin(16))/\Z_2$ which is not
simply connected. If we are willing to accept $(Spin(16) \times
Spin(16))/\Z_2$ as the 2 dimensional symmetry group, it turns out to
be incompatible with the symmetry in 3 dimensions. If the gravitini
are in the $\mathbf{(16,1) \oplus (1,16)}$ of $(Spin(16) \times
Spin(16))/\Z_2$, as suggested in \cite{Duff:2003ec}, then the 3-d
symmetry-group with $so(16)$ algebra must be
embedded diagonally in $(Spin(16) \times Spin(16))/\Z_2$ (because
both chiralities of the 2-d gravitini must give rise to the
$\mathbf{16}$ of the 3-d gravitini). But the diagonal
group in $(Spin(16) \times Spin(16))/\Z_2$ is $SO(16)$, as a
computation demonstrates.  A set of representatives generating all congruence
classes of $(Spin(16) \times Spin(16))/\Z_2$ is $\mathbf{(16,1)}$,
$\mathbf{(128_s,128_s)}$ and $\mathbf{(128_s,128_c)}$. These decompose as:
\bea
\mathbf{(16,1)} & \rightarrow & \mathbf{16} \non \\
\mathbf{(128_s,128_s)} & \rightarrow & \mathbf{1 \oplus 120 \oplus
  1820 \oplus 8008 \oplus 6435} \\
\mathbf{(128_s,128_c)} & \rightarrow & \mathbf{16 \oplus 560 \oplus
  4368 \oplus 11440 } \non
\eea
which are all $SO(16)$ irreps. Hence also assuming $Spin(32)$ as a
symmetry-group appears to be inconsistent with the established
higher dimensional symmetry groups, at least with the embeddings
suggested in \cite{Duff:2003ec}. 

We cannot find an objection against $\widetilde{H}_2=Spin(16) \times
  Spin(16)$ this way, but it seems appropriate to conclude here with
  the observation that the caveats mentioned by Duff and Liu
  \cite{Duff:2003ec} are very real, and seem fatal in some cases.   

\section{Concluding remarks} 

In this article we have tried to draw attention to the often neglected
global properties of Lie-groups and their subgroups, as they appear in
supergravity and related theories. To demonstrate that this is not
merely a mathematical enterprise, but actually can give decisive tools,
we have illustrated our methods by two applications.

First we have discussed how the inclusion of fermions in theories that
give rise to cosets $G/H$ in 3 dimensions leads to restrictions on
the topology of the compact subgroups $H$. Therefore, the first
fundamental group $\pi_1(H)$ of $H$ indicates whether
the theory can be interpreted as a reduction of a higher dimensional
theory.  

In a second application, we have studied the actual topology of the
compact subgroups $H_d$ of the Cremmer-Julia groups $G_d$ of maximal
supergravity. The semi-simple groups appearing here are
two-fold connected (for $d < 8$). Additional fermions transform in the double cover
$\widetilde{H_d}$ of $H_d$, and hence do \emph{not} form representations of
$H_d$, and can\emph{not}, even with the inclusion of additional auxiliary
fields, be promoted to representations of the Cremmer-Julia groups
$G_d$. In particular, this identifies the symmetry group of 3
dimensional supergravity as $Spin(16)$ (and not $SO(16)$ or
$Spin(16)/\Z_2$). 

We then showed that the recently proposed $SL(32,\R)$
\cite{Hull:2003mf} does not contain $Spin(16)$; it does contain
$SO(16)$ but this is not the proper symmetry group of 3 dimensional
supergravity. This demonstrates that
there are symmetries of supergravity not contained in $SL(32,\R)$. The
group $\widetilde{H}_1=Spin(32)$ does have $Spin(16)$ subgroups, but
their embedding seems incompatible with the representations of the
gravitini; an embedding compatible with the gravitini representations
gives $SO(16)$ and not $Spin(16)$. Hence $SL(32,\R)$ and
$Spin(32)$ seem to be ruled out as candidate symmetries of 11
dimensional supergravity, and M-theory, at least as local symmetries
in formulations of the type presented in \cite{deWit:1985iy}. 

This does not mean of course that the idea of ``generalized
holonomy'' is wrong, but indicates that the concept is much more
involved. As 11 dimensional Lorentz-spinors are real and have 32 components, 
$GL(32,\R)$ or $SL(32,\R)$ and their subgroups may seem natural guesses for
symmetry groups of M-theory. Indeed, $SL(32,\R)$ acts as an
automorphism on the supersymmetry algebra but this is not sufficient to
elevate it to a symmetry of the theory\footnote{The proposal that one may enlarge
  the tangent space group to a generalized structure group if the
  Killing spinor equation has hidden symmetries \cite{Duff:2003ec} is
  essentially a rephrasing of this, and the same criticism applies:
  Hidden symmetries of the Killing spinor equations are a necessary,
  but far from sufficient ingredient for enlarging the structure
  group. Note that the supercovariant derivative used in
  \cite{Duff:2003ec, Hull:2003mf, Papadopoulos:2003pf} is not
  $SL(32,\R)$ covariant.}; one also has to require that the symmetry is properly
represented on (all fields in) the theory. Even then, this is still not
sufficient to elevate it to a \emph{local} symmetry.

It is interesting to compare with the situation for supergravity in 3
dimensions, which played an important part in our story. There, the spinor parameters have a
symmetry group $SO(16)$, they are real and come in 16 dimensional
irreps. That does not imply however that $SO(16)$ is a symmetry of the
theory; the bosons and fermions of 3-d supergravity transform in
irreps of $Spin(16)$, that are not irreps of $SO(16)$. Another
counterexample to the assertion ``automorphism of symmetry
algebra implies symmetry of the theory'' is given by
a theory with a symmetry group $G$ allowing an outer
automorphism $T$. Such a theory need not be invariant under the outer
automorphism, because nothing guarantees that the irreps of $G$ fill
out multiplets of $G \ltimes T$.

The group $SL(32,\R)$  was also suggested in \cite{Barwald:1999is} as
a possible symmetry group of M-theory, and identified
\cite{West:2003fc} with the $SL(32,\R)$ of \cite{Hull:2003mf}. This claim is due to the
identification of parts of the symmetry of $SL(32,\R)$. Since it is
believed that M-theory has large symmetry algebras, there seems to be
no problem in promoting these symmetries to be substructures of
some bigger group/algebra.

That the ``natural'' symmetry groups $SL(32,\R)$ and $GL(32,\R)$ do
not contain all known symmetries of 11 dimensional supergravity is an
indication that we might have to transcend beyond Lorentz-covariant
formalisms, and turn to bigger finite, or infinite groups (with the
immediate question how to realize them on a 32 component spinor).
This brings the infinite-dimensional groups $E_{10}$ and $E_{11}$ (and
  their subgroups) back into the picture
  \cite{Julia:1982gx,West:2003fc, E10E11}. Note however, that in the
  light of the previous discussion it seems unavoidable that also these groups can
  at most appear in the numerator of some quotient, and that fermions
  in the theory should appear in representations of a covering of the
  quotient group in the denominator. This in turn seems to imply that
  not $E_{11}$, but at most the ``simply connected covering'' of its
  ``maximal compact subgroup'' can give a symmetry of
  M-theory. However, in absence of a more detailed understanding of
  these groups, it is not even clear what the precise meaning of the phrases in
  parentheses (which were written by analogy to finite-dimensional groups) is.

{\bf Acknowledgements} I am extremely thankful to Hermann Nicolai and
Chris Hull for discussions and extensive correspondence. The author is
supported in part by the ``FWO-Vlaanderen'' through project G.0034.02,
in part by the Federal office for Scientific, Technical and Cultural
Affairs trough the Interuniversity Attraction Pole P5/27 and in part
by the European Commision RTN programme HPRN-CT-2000-00131, in which
the author is associated to the University of Leuven.

\appendix
\section{Compact subgroups and outer automorphisms} \label{app}

Theorem 1 and its applications mentioned in section \ref{topo} cover
all the cases where the non-compact form of $G$ is generated by an
inner automorphism. The other case, where the non-compact form is
generated by an outer automorphism will be treated in this appendix. 

Only the groups corresponding to the Dynkin diagrams of $A_n$ ($n > 1$),
$D_n$ ($n \geq 3$), and $E_6$ allow outer automorphisms. We discuss
these case by case.

\subsection{$A_n$}

$sl(n,\R), (n>2)$: The maximal compact subgroup is $SO(n)$. The
  embedding proceeds via the fundamental $\mathbf{n}$ irrep of $SL(n,\R)$,
  that descends to the $\mathbf{n}$ vector irrep of $SO(n)$. All irreps of
  $SL(n,\R)$ can be found by tensoring the fundamental $\mathbf{n}$
  with copies of itself and (anti-)symmetrizing. The irreps of $SO(n)$
  as embedded in $SL(n,\R)$ are found in the same way. This procedure
  never gives any of the spin irreps of $Spin(n)$, and hence
  $\pi_1(SO(n)) = \Z_2$. 

$su^{*}(2n)$: This has maximal compact subgroup $Sp(n)$. The embedding
  proceeds via the fundamental $\mathbf{2n}$ irrep of $SU^*(2n)$ that
  descends to the fundamental of $Sp(n)$. Again all other irreps can
  be found by tensoring. The fundamental of $Sp(n)$ realizes the full
  center, hence we are dealing with $Sp(n)$ and not $Sp(n)/\Z_2$.

\subsection{$D_n$}

$so(1,n)$: This real form is generated by an outer
  automorphism for $n$ odd. Three representatives generating all congruence
  classes are the vector irrep $\mathbf{(n+1)}$ and the
  spin irreps $\mathbf{(2^{\frac{n-1}{2}})_{s,c}}$. These decompose to
  the compact subgroup $Spin(n)$ as
\be
\mathbf{(n+1)} \rightarrow \mathbf{n \oplus 1} \qquad
\mathbf{(2^{\frac{n-1}{2}})_{s,c}} \rightarrow 
\mathbf{2^{\frac{n-1}{2}}}
\ee
The presence of the spin irrep of $Spin(n)$ implies that the subgroup is simply connected.

$so(p,q)$: This real form is generated by an outer automorphism if
$p$ and $q$ are \emph{both} odd. Taking again the vector and spin
irreps of as representatives, decompositions into the $Spin(p) \times
Spin(q)$ subgroup are
\be
\mathbf{(p+q)} \rightarrow \mathbf{(p,1) \oplus (1,q)} \qquad
\mathbf{(2^{\frac{p+q}{2}-1})_{s,c}} \rightarrow
\mathbf{(2^{\frac{p-1}{2}},2^{\frac{q-1}{2}})} 
\ee
These irreps do not realize the diagonal $\Z_2$ in the $\Z_2 \times \Z_2$
center of $Spin(p) \times Spin(q)$, hence the proper subgroup is
$(Spin(p) \times Spin(q))/\Z_2$. 

\subsection{$E_6$}

$E_{6(6)}$: The maximal compact subgroup is $Sp(4)$. A possible
  representative is the $\mathbf{27}$ dimensional irrep of $E_{6(6)}$
  from which all other irreps can be generated by tensoring. But this
  irrep descends to the $\mathbf{27}$, which is an irrep of
  $Sp(4)/\Z_2$. Hence, the compact subgroup is a two-fold connected
  group.

$E_{6(-26)}$: The compact subgroup is $F_4$. The universal cover of
  $F_4$ has a trivial center, and hence all possible forms of the
  compact group $F_4$ are simply connected.

\end{document}